\documentclass[twocolumn,showpacs,preprintnumbers,amsmath,amssymb, prl, reprint]{revtex4-1}

\usepackage{stmaryrd}
\usepackage{txfonts}
\usepackage{amssymb}
\usepackage{mathrsfs}
\usepackage{graphicx}
\usepackage{dcolumn}
\usepackage{bm}
\usepackage{epsfig}
\usepackage{color}                    
\usepackage{hyperref}                 

\begin{document}
\preprint{\href{http://dx.doi.org/10.1103/PhysRevLett.110.207202}{S.-Z. Lin, C. Reichhardt, C. D. Batista and A. Saxena , Phys. Rev. Lett. {\bf 110}, 207202 (2013).}}

\title{Driven skyrmions and dynamical transitions in chiral magnets}

\author{Shi-Zeng Lin}
\affiliation{Theoretical Division, Los Alamos National Laboratory, Los Alamos, New Mexico 87545, USA}

\author{ Charles Reichhardt}
\affiliation{Theoretical Division, Los Alamos National Laboratory, Los Alamos, New Mexico 87545, USA}

\author{Cristian D. Batista}
\affiliation{Theoretical Division, Los Alamos National Laboratory, Los Alamos, New Mexico 87545, USA}

\author{Avadh Saxena}
\affiliation{Theoretical Division, Los Alamos National Laboratory, Los Alamos, New Mexico 87545, USA}

\begin{abstract}
We study the dynamics of skyrmions in chiral magnets in the presence of a spin polarized current. The motion of skyrmions in the ferromagnetic background excites spin waves and contributes to additional damping. At a large current, the spin wave spectrum becomes gapless and skyrmions are created dynamically from the ferromagnetic state. At an even higher current, these skyrmions are strongly deformed due to the damping and become unstable at a threshold current, leading to a chiral liquid. We show how skyrmions can be created by increasing the current in the magnetic spiral state. We then construct a dynamic phase diagram for a chiral magnet with a current. The instability transitions between different states can be observed as experimentally clear signatures in the transport measurements, such as jumps and hysteresis.  
 \end{abstract}
\date{\today}
\maketitle

\noindent {\it Introduction --} Skyrmions as topological objects were first proposed as a model for baryons \cite{Skyrme61} 
and were first realized experimentally in condensed matter systems in quantum-Hall ferromagnets. \cite{Barrett95} 
Skyrmions were also predicted to be stable in chiral magnets, where the inversion symmetry is broken due to the spin-orbit coupling.\cite{Bogdanov89,Bogdanov94,Rosler2006} In this context, the spins wrap a sphere when one moves from the center of the skyrmion to infinity. 
Hexagonal skyrmion lattices were observed in the $A$ phase of a metallic ferromagnet MnSi \cite{Muhlbauer2009} as well as in compounds with a B20 structure \cite{Munzer10,Pfleiderer10}. More recently skyrmion lattices have been directly imaged 
in thin films of $\rm{Fe_{0.5}Co_{0.5}Si}$ by Lorentz force microscopy \cite{Yu2010a,Yu2011}. The continuing identification of more materials with skyrmion lattices suggests that they are ubiquitous in magnetic metals without inversion symmetry.   

One of the main themes in the field of spintronics is 
the manipulation of spin textures by electric currents. The main player is 
magnetic domain walls (DWs) and their controlled motion by current is  under active study. It has been shown \cite{Yamanouchi2004} that the threshold current to move DWs is extremely high, on the order of $10^5 - 10^6\ \rm{A/cm^2}$, which is detrimental for applications due to the intensive Joule heating.
Skyrmions that can also be driven by a current are promising candidates for the application of spintronics as information carriers, but their potential use for this purpose depends crucially on both controlled creation of skyrmions and manipulation 
of their motion. Recently it was demonstrated experimentally that the threshold current to move a skyrmion is 4 to 5 orders of magnitude smaller than that required to move DWs \cite{Jonietz2010,Yu2012,Schulz2012}, indicating
that skyrmions could have a tremendous advantage over DWs since the Joule heating can be reduced significantly. 
Thus, an understanding of the dynamics of skyrmions is both necessary and timely, particularly  the stability of the driven skyrmions and how fast they can
be manipulated.

In this Letter, we study the motion of a single skyrmion in a ferromagnetic (FM) background and the flow of a skyrmion lattice in the presence of a current. Due to the damping, the skyrmion acquires a velocity transverse to the current. The motion of the skyrmion perturbs the FM background
causing the radiation of a spin wave. This radiation is enhanced when the velocity of skyrmion reaches a threshold 
and the spin wave spectrum becomes gapless, leading  to 
an instability where multiple skyrmions are created and the voltage, 
which is proportional to the number of skyrmions, increases sharply.
At a higher current, the skyrmion structure is strongly distorted due to the damping and finally the skyrmions become unstable, resulting in a chiral liquid with random and chaotic precession of spins. The ratio between the longitudinal and transverse voltages jumps at the instability.
We also show that the skyrmion lattice can be dynamically created from the spiral state under current. Based on these results, a dynamic phase diagram for chiral magnets with a current is constructed. The 
onset of the different instability transitions can be observed from transport measurements in experiments, including pronounced jumps and hysteresis effects
in the current-voltage curves.

\noindent {\it Model --}
We consider a thin film of chiral magnets, which is described by the spin Hamiltonian with the Dzyaloshinskii-Moriya (DM) interaction in a dimensionless form \cite{Bogdanov89,Bogdanov94,Rosler2006,Han10,Rossler2011},
\begin{equation}\label{eq1}
\mathcal{H}=\int d\mathbf{r}^2 \left[\frac{J_{\rm{ex}}}{2}(\nabla \mathbf{n})^2+D\mathbf{n}\cdot\nabla\times \mathbf{n}-\mathbf{H}_a\cdot\mathbf{n} \right],
\end{equation} 
where $J_{\rm{ex}}$ is the exchange constant, $D$ is the DM interaction  due to spin-orbit coupling \cite{Dzyaloshinsky1958,Moriya60,Moriya60b}, and the last term accounts for the Zeeman interaction. Here $\mathbf{n}(x, y)$ is a unit vector describing the direction of magnetic moment. The external field is perpendicular to the film $\mathbf{H}_a=H_a\hat{z}$. The phase diagram of chiral magnets obeying Eq. \eqref{eq1} at temperature $T=0$\ \rm{K} is summarized as follows. \cite{Rossler2011} At $H_a=0$, the system favors a spiral configuration of magnetic moments. Upon increasing $H_a$, the hexagonal skyrmion lattice is stabilized at $H_a> 0.2 D^2/J_{\rm{ex}}$. At higher fields $H_a>0.8 D^2/J_{\rm{ex}}$, the ferromagnet becomes the ground state. Both transitions are of the first order. 

In the presence of a current, the localized moment experiences a spin transfer torque due to the conduction electrons. The dynamics of the magnetic moments is described by the Landau-Lifshitz-Gilbert equation \cite{Bazaliy98,Li04,Tatara2008}
\begin{equation}\label{eq2}
{\partial _t}{\bf{n}} = ({{\bf{J}} }\cdot\nabla) {\bf{n}} - \gamma {\bf{n}} \times {{\bf{H}}_{\rm{eff}}} + \alpha {\partial _t}{\bf{n}} \times {\bf{n}},
\end{equation}
where the first term on the right hand side (r.h.s.) represents the spin-transfer torque with current density $\bf{J}$ and the last r.h.s. term is the Gilbert damping. \footnote{There is additional damping due to the electric fields induced by the motion of skyrmions \cite{Zang11}, which can be taken into account by using an effective damping coefficient.} The effective field is $\mathbf{H}_{\rm{eff}}=-\delta \mathcal{H}/\delta {\bf{n}}=J_{\rm{ex}}{\nabla ^2}{\bf{n}} - 2D\nabla  \times {\bf{n}} + {\bf{H}_a}$. We solve Eq. \eqref{eq2} numerically \cite{noteNumerics} to obtain the current-voltage (I-V) characteristics with the electric field given by ${\bf{E}} ={\bf{n}}\cdot(\nabla {\bf{n}} \times {\partial _t}{\bf{n}})$. \cite{Zang11} The total number of skyrmions in the film is $Q=\int d\mathbf{r}^2q(\mathbf{r})$ with the skyrmion density $q(\mathbf{r}) = {\bf{n}}\cdot({\partial _x}{\bf{n}} \times {\partial _y}{\bf{n}})/(4\pi)$. \cite{SimonsQFT}

\begin{figure}[t]
\psfig{figure=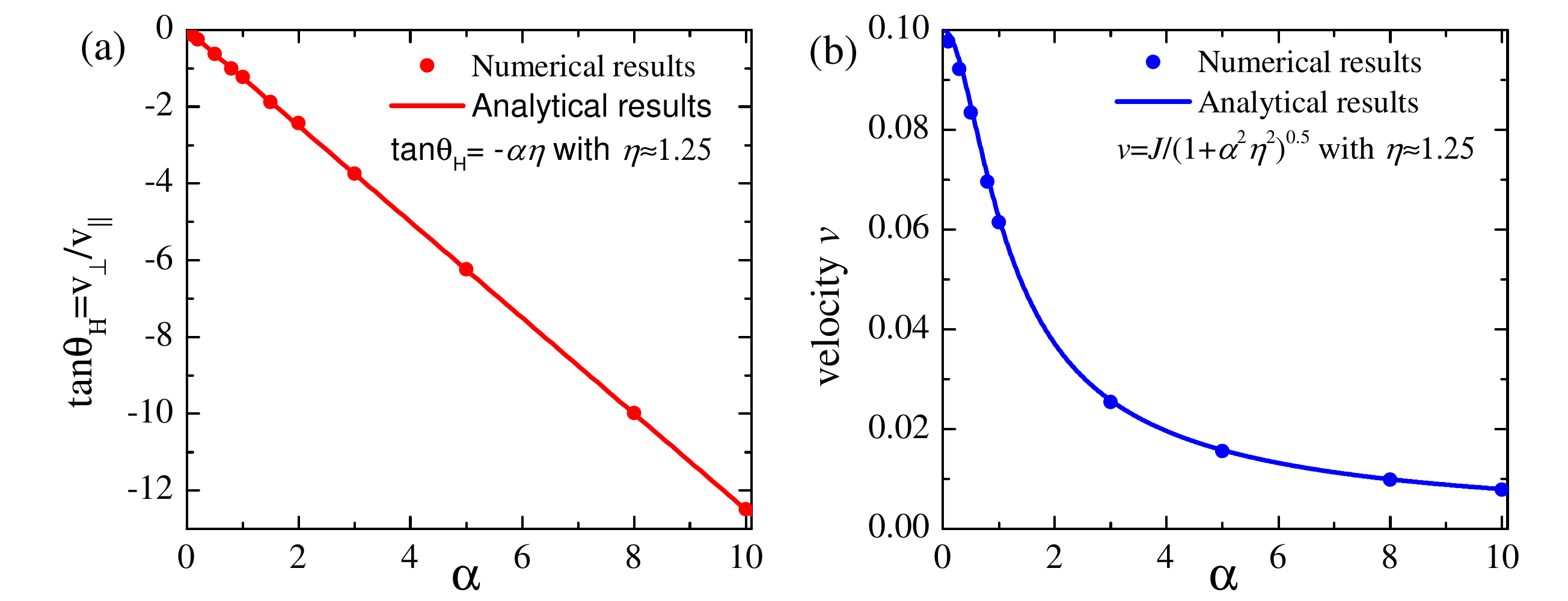,width=\columnwidth}
\caption{\label{f1}(color online) (a) The Hall angle $\tan\theta_H$ and (b) the velocity $v$ vs $\alpha$ for the motion of a single skyrmion obtained numerically (dots) and analytically (lines) at $J=0.1$.}
\end{figure}

\noindent {\it Dynamics of a single skyrmion --} Let us first consider the motion of a single skyrmion. In the absence of damping $\alpha=0$, Eq. \eqref{eq2} has an exact solution 
$\mathbf{n}_S(\mathbf{r}- \mathbf{v} t)$ with $\mathbf{v}=-\mathbf{J}$, where $\mathbf{n}_S(\mathbf{r})$ is the structure of the skyrmion in the static case $J=0$. This solution corresponds to the comoving of skyrmion and conduction electrons, therefore the spin-transfer torque is zero.  With damping $\alpha>0$, the structure of the skyrmion deforms. For low velocities, we can use the quasi-static approximation that the structure of the skyrmion is the same as that in the static case. The velocity of the skyrmion is given by \cite{Zang11} 
\begin{equation}\label{eq3}
\mathbf{v} =  - {\bf{J}} - \alpha \eta {\bf{\hat z}} \times \mathbf{v}
\end{equation}
with $\eta=\eta_{\mu}=4\pi {[\int {{d}}{\mathbf{r}^2}{({\partial _{\mu}}{\bf{n}})^2}]^{ - 1}}$ and $\mu=x,\ y$. The skyrmion acquires a velocity component perpendicular to the current due to the damping. The velocity parallel to the current is $v_{\parallel}=-J/({1+\alpha^2\eta^2})$ and the perpendicular velocity is $v_{\perp}=J\alpha\eta/({1+\alpha^2\eta^2})$. The corresponding voltages across the skyrmion are $V_{\parallel}=-4\pi v_{\perp}$ and $V_{\perp}=4\pi v_{\parallel}$. Thus the Hall angle is $\theta_H\equiv \tan^{-1}(v_{\perp}/v_{\parallel})=-\tan^{-1}(\alpha\eta)$.  We calculate the Hall angle and velocity of a single skyrmion numerically. The velocity is defined as $\mathbf{v}=\dot{\mathbf{s}}(t)$ with $\mathbf{s}$ the center of mass $\mathbf{s}=\frac{1}{Q}\int d \mathbf{r}^2 q(\mathbf{r})\mathbf{r}$. At a small current (velocity), the numerical results of Hall angle and velocity are consistent with the analytical calculations, as shown in Fig. \ref{f1}. 

\begin{figure*}[t]
\psfig{figure=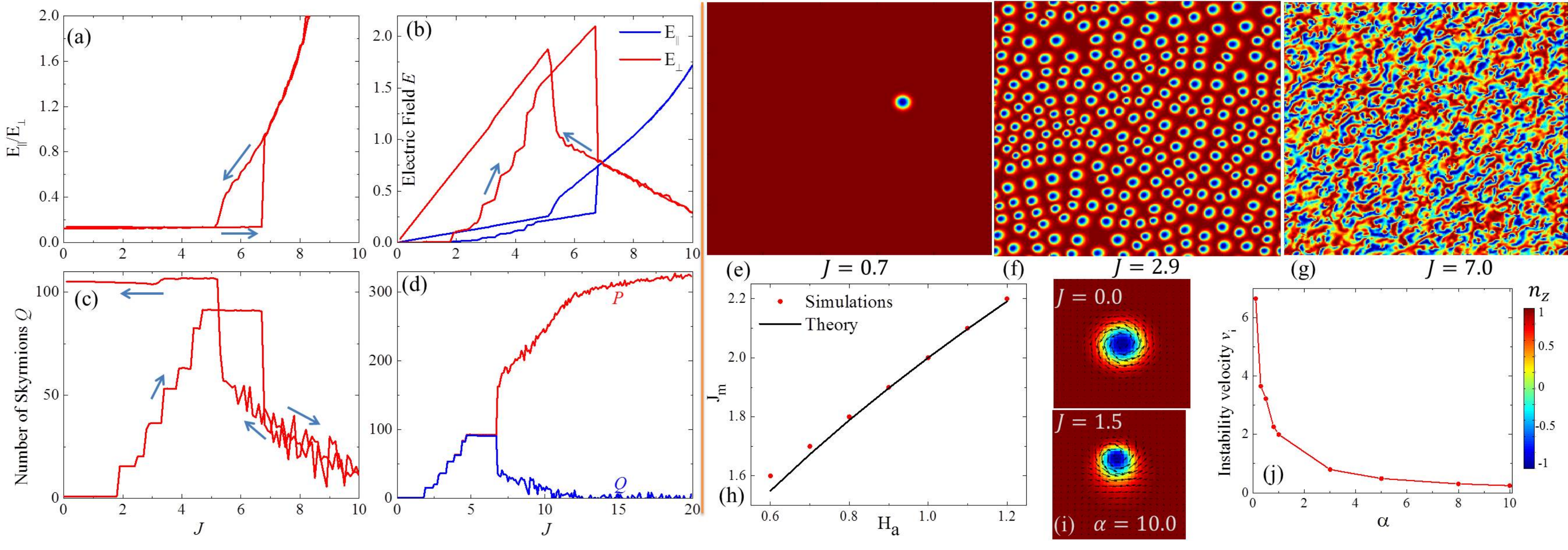,width=18cm}
\caption{\label{f2}(color online) 
(a) Ratio of the electric field parallel to the current $E_{\parallel}$ to the electric field perpendicular to the current $E_{\perp}$, (b) electric field $E$, (c) total number of skyrmions $Q$, (d) chirality $P\equiv\int d\mathbf{r}^2|q(\mathbf{r})|$ and $Q$ vs the spin current $J$. The arrows in (a-c) indicate the direction of current sweep while in (d) only the results with increasing current are shown. (e-g) Configurations of $n_z$ in the case of (e) a single skyrmion, 
(f) multiple skyrmions, and (g) the chiral liquid phase at different currents. Here $\alpha=0.1$ and $H_a=0.6$. (h) Comparison between analytical (line) and numerical (dots) results for the current $J_m$ at which the instability occurs and multiple skyrmions are created. (i) Comparison between spatial structure of a skyrmion at rest and at a large current, showing distortion. The vectors in the plots denote the $n_x$ and $n_y$ components, and color denotes the $n_z$ component. (j) The velocity $v_i$ where skyrmions become unstable vs $\alpha$.}
\end{figure*}

\begin{figure}[b]
\psfig{figure=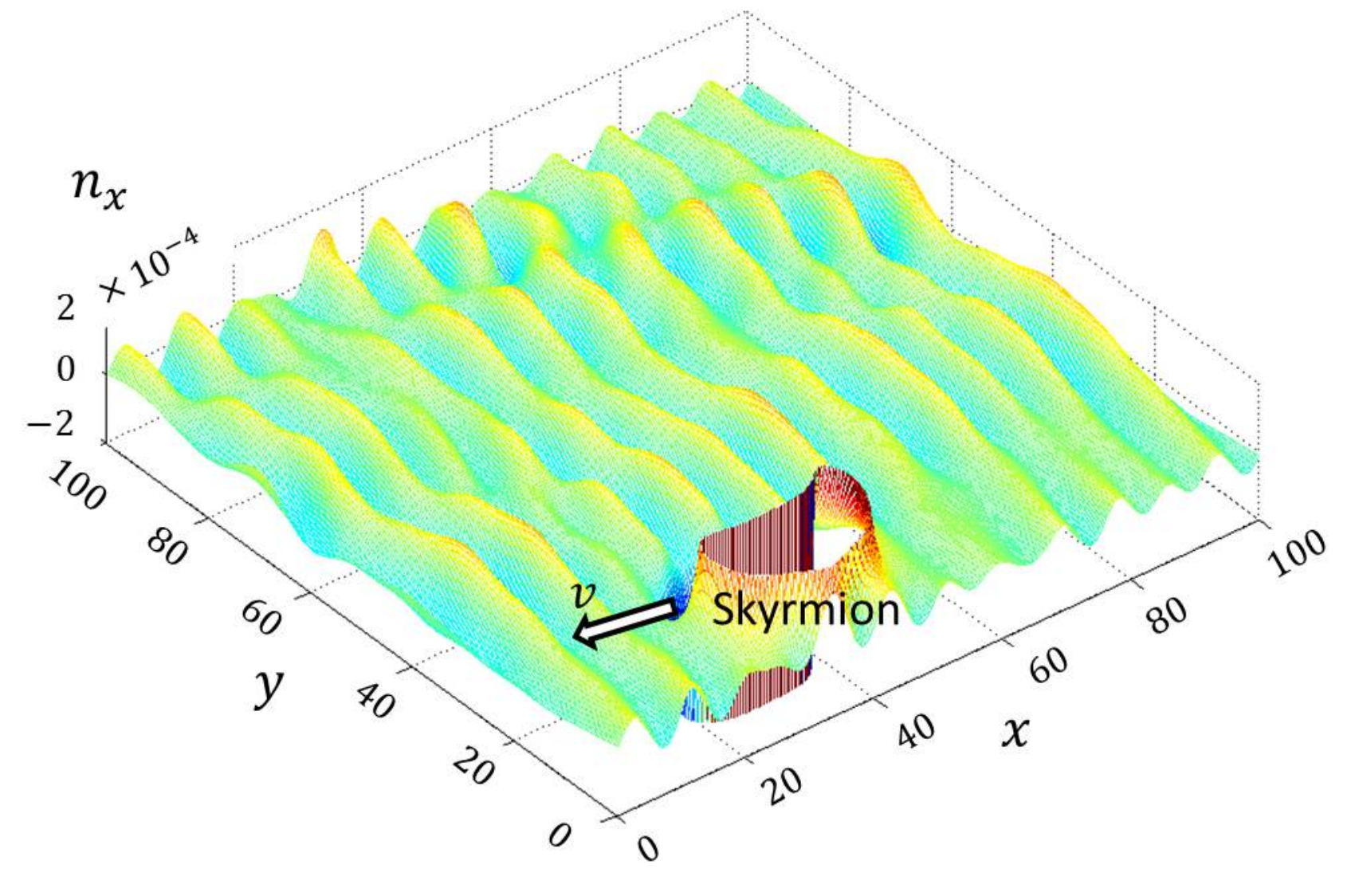,width=\columnwidth}
\caption{\label{f3}(color online) Spatial structure of the 
$n_x$ component of the spin wave radiated due to the motion of a skyrmion. 
The skyrmion location is labeled and the arrow denotes the direction of skyrmion motion. Here $\alpha=0.1$, $J=1.4$ and $H_a=0.6$.}
\end{figure}

When the current (velocity) increases, the skyrmion starts to deform (for large $\alpha$), and thus the quasi-static approximation in Eq. \eqref{eq3} becomes questionable. We calculate numerically the I-V curve with a single skyrmion at $J=0$ as an initial state. The resulting I-V curve and dependence of $Q$ on $J$ are shown in Figs. \ref{f2} (a-c). For a small current, the motion of the skyrmion is stable, as shown in Fig. \ref{f2}(e). The movement of the skyrmion perturbs the ferromagnetic background and excites spin waves. The effect of the radiation of a spin wave is more prominent when the skyrmion is accelerated or decelerated by a time-dependent current or scattered by impurities. For a constant motion of the skyrmion, maximal radiation occurs when the resonance condition $\Omega(\mathbf{k})=\mathbf{v}\cdot\mathbf{k}$ is satisfied, where 
\begin{equation}\label{eq6}
\Omega(\mathbf{k})  = \mathbf{J}\cdot \mathbf{k}+ \frac{\gamma (1+i\alpha) }{\alpha ^2+1}\left(H_a+ J_{\rm{ex}} \mathbf{k}^2\right),
\end{equation}
is the spin wave spectrum determined from Eqs. \eqref{eq1} and \eqref{eq2} taking the dissipation into account. It requires a threshold current $J_r$ above which the resonance condition $\Omega(\mathbf{k})=\mathbf{v}\cdot\mathbf{k}$ can be satisfied. For weak damping $\alpha\ll 1$ where $\mathbf{v}\approx -\mathbf{J}$, $J_r=\gamma  \sqrt{H_a J_{\text{ex}}}/(\alpha ^2+1)$. The dominant wave vector is $k_x=\sqrt{H_a/J_{\rm{ex}}}$ and $k_y=0$. One typical configuration of $n_x$ is illustrated in Fig. \ref{f3} which clearly shows the radiation of the spin wave due to the motion of the skyrmion. For strong damping, the excited spin wave decays quickly  because the amplitude of the spin wave decreases with the damping constant $\alpha$.  The radiation of spin waves thus gives an additional contribution to the damping of skyrmion. According to the simulations, this contribution is small because the amplitude of the spin wave is weak. The radiation of collective excitations due to the motion of driven topological objects is well known in other systems, such as Josephson junctions \cite{HuReview09} and DWs \cite{Bouzidi90,Wieser10}. 

\noindent {\it Dynamical creation and destruction of skyrmions--} The spin wave excitation spectrum $\Omega(\mathbf{k})$ [Eq. \eqref{eq6}] becomes gapless when the applied current reaches the value $J_m={2 \gamma  \sqrt{H_a J_{\text{ex}}}}/({\alpha ^2+1})$.  This gapless condition signals an instability of the FM state into a helical state whose $xy$-spin component has spiral correlations with wave vector ${\mathbf {k}}_m = (k_x=\sqrt{H_a/J_{\rm{ex}}}, k_y=0)$. Simulations show that the amplitude of the spiral $xy$-component grows until the partially  saturated FM state is destroyed via proliferation of multiple skyrmions  [see Figs. \ref{f2} (c) and \ref{f2} (f)]. The presence of a moving skyrmion affects the relaxation of the original FM state into the skyrmion-rich state in the neighborhood of the moving skyrmion because spin fluctuations are locally enhanced via radiation of spin waves.  The amplitude of the ${\mathbf{k}} = {\mathbf{k}}_m$ spin wave  keeps increasing  until more skyrmions are created around the original moving skyrmion. The new skyrmions move in response to the applied current and radiate spin waves that create a new generation of skyrmions. This multiplicative process repeats, until the system is filled with a finite concentration of skyrmions. As more and more energy is pumped into the system, the density of skyrmions keeps increasing and saturates around a value corresponding to the equilibrium state, see Fig. \ref{f2} (c).

Skyrmions are created after the instability of the FM state because the current density, $\mathbf{J}$, couples to the emergent vector potential $ \mathbf{A}\equiv {i c \hbar  b^{\dagger } \nabla b}/{e}$  generated by non-coplanar spin configurations via the Lagrangian term $\mathcal{L}_{JA}=\mathbf{J}\cdot\mathbf{A}$ (where $b$ is the spin coherent state \cite{SimonsQFT}). This coupling stabilizes spin states with non-zero $\mathbf{A}$. In the presence of the DM interaction, the lowest energy state with non-zero $\mathbf{A}$ is a state with skyrmions. The  electric field  increases in a series of steps [Fig.~\ref{f2} (b)] because it is proportional to the number of skyrmions $Q$ that increases in the same way [see Fig. \ref{f2} (c)]. The change of current does not modify the direction of motion  because the ratio $E_{\parallel}/E_{\perp}$ is independent of $J$ [Fig. \ref{f2} (a)].  Figure \ref{f2} (h) shows that our analytical expression for $J_m$ reproduces the current value that is obtained from our simulations (value required to trigger the proliferation of  skyrmions) in the weak damping regime, $\alpha\ll 1$, that is relevant for most materials. For strong damping, $\alpha>1$, the energy pumped  into the spin waves is quickly damped and  $J_m$ shifts to values that are higher than the one  predicted by the linear analysis.

\begin{figure*}[t]
\psfig{figure=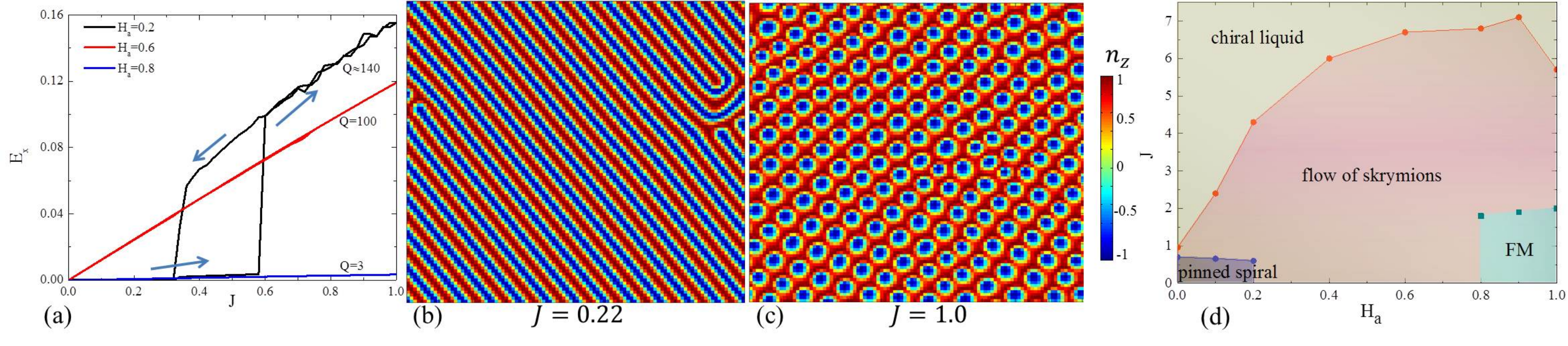,width=18cm}
\caption{\label{f4}(color online) (a) Dependence of the longitudinal electric field $E_x$ on spin current at different fields $H_a=0.2$, $H_a=0.6$ and $H_a=0.8$. For $H_a=0.2$, the number of skyrmions $Q$ is not conserved at high currents and only an averaged value of $Q$ is given. Arrows denote the direction of current sweep. (b-c) Configurations $n_z$ for (b) the spiral phase and (c) the skyrmion lattice obtained with $H_a=0.2$. The spiral phase is not perfect 
and contains some defects (right part of the plot) after the annealing process. (d) Dynamic phase diagram for $J$ vs $H_a$ of a chiral magnet 
obeying the Hamiltonian in Eq. \eqref{eq1} in the presence of 
a spin current. Here $\alpha=0.1$.}
\end{figure*}

The skyrmion is no longer a circular object under large current and 
develops a tail opposite to the direction of motion, as shown in Fig. \ref{f2} (i). The distortion is stronger for larger $\alpha$. At an even higher current, the structure of the skyrmion is so strongly distorted that the skyrmion becomes unstable.  Here the relation $E_{\perp}/E_{\parallel}=-\alpha\eta$ breaks down as shown in Fig. \ref{f2} (a), 
indicating that there is an upper velocity limit for skyrmion motion due to the distortion induced by damping. When the current is reduced, the system recovers to a skyrmion phase at a current smaller than that at which the skyrmion was lost in the ramping up process, indicating hysteresis. As the current is further reduced, the number of skyrmions remains constant down to $J=0$. The velocity $v_i$ where the skyrmions become unstable decreases with the damping constant $\alpha$ as $v_i\sim 1/\sqrt{1+\alpha^2\eta^2}$, as shown in Fig. \ref{f2}(j). The distortion mediated instability of a topological object has been demonstrated in other systems, such as Larkin-Ovchinnikov instability of the vortex lattice in superconductors \cite{Larkin65} and Walker breakdown of DWs \cite{Schryer1974}.

The disordered state shown in Fig.~\ref{f2} (g) is a chiral liquid, as revealed by a numerical calculation of $Q$ and $P\equiv\int d\mathbf{r}^2|q(\mathbf{r})|$ [see Fig. \ref{f2} (d)]. $Q\rightarrow 0$ while $P$ increases and then saturates as a function of increasing current. This behavior indicates that the scalar chirality $q(\mathbf{r})$ (or emergent magnetic field)  has strong real space fluctuations that change its sign and the average chirality is much smaller than the amplitude of the fluctuations. The sawtooth-like curve shown in Fig.~\ref{f2} (c) for the high current region indicates that the average chirality also fluctuates as a function of increasing current. This chiral liquid can be treated as a superposition of skyrmion and anti-skyrmion liquids. According to Eq. \eqref{eq3}, the electric field parallel to the current, $E_{\parallel}$, does not depend on the chirality sign. In contrast, the electric field perpendicular to the current, $E_{\perp}$, changes sign when the chirality is reversed. Thus, $E_{\parallel}$ increases with $J$ while $E_{\perp}$ decreases in the chiral liquid phase [see Fig.~\ref{f2} (b)]. Meanwhile $E_{\perp}$ fluctuates strongly because it is proportional to the small value of the average chirality, while $E_{\parallel}$ exhibits a much smoother behavior because it is determined by $P$. Both behaviors are consistent with the numerical results shown in Fig.~\ref{f2} (b). We remark that these dynamic phase transitions show jumps in the I-V curves, implying that they can be observed experimentally by transport measurements. In this case, one needs to subtract the usual contribution from the electronic background.

\noindent {\it Dynamics Starting from the Ground State Configurations--} 
In order to understand how the creation and destruction of the current driven states could be observed experimentally, we next consider the dynamics starting from the ground state configurations, which are prepared carefully by numerical annealing. In the ferromagnetic phase, the effect of the current vanishes since $\nabla\cdot \mathbf{n}=0$ in this case, so that $\mathbf{E}=0$.  With fluctuations, the current couples with the fluctuations of magnetization and creates skyrmions at $J_m$ where the spectrum of spin waves becomes gapless. The I-V curves for the spiral and skyrmion phases are shown in Fig. \ref{f4}(a). For a small current, the spiral structure does not move and no voltage is induced in the system, as illustrated in Fig. \ref{f4} (b). 
This intrinsic pinning of the spiral structure is due to the DM interaction.
Let us consider a spiral structure along the $x$ direction $\mathbf{n}=(0,\ \cos(q x),\ \sin(q x))$, where $q$ is the wave number of the spiral structure. The spin-transfer torque is $J_x q(0,\ -\sin(q x),\ \cos(q x))=-\gamma \mathbf{n}\times\mathbf{H}_s$ with $\mathbf{H}_s=J_x q\hat{x}/\gamma$. Thus the effect of the spin transfer torque is equivalent to a magnetic field along the $x$ direction. This field tends to align the spin along the $x$ direction due to the Gilbert damping term, which costs energy as a result of the DM interaction. Thus, intrinsic pinning of skyrmions exists in clean systems \cite{Iwasaki2013}, similar to the case of DWs \cite{Tatara04}.

As the current is increased, the spiral structure becomes unstable and a lattice of skyrmions emerges above a critical current, as shown in Fig. \ref{f4} (c). At $J\approx 0.34$, a small number of skyrmions ($Q=5$) is created while most of the system retains the spiral structure. At $J\approx 0.6$, the remaining spiral structure is converted into the skyrmion lattice. These skyrmions move freely under an applied current, which induces transverse and longitudinal electric fields. The electric fields are proportional to the number of skyrmions, so they decrease with increasing $H_a$ because the skyrmion density decreases. When the current is increased further, the skyrmions become unstable and the system evolves into the chiral liquid phase discussed above. At low fields where the spiral structure is the ground state, hysteresis occurs upon decreasing the current, while at high fields where the skyrmion lattice is the ground state, no hysteresis is observed. From these results, we construct a dynamic phase diagram for chiral magnets in a magnetic field under spin polarized current, as shown in Fig. \ref{f4} (d).  There is hysteresis when the current is swept across the phase boundary, so we show the location of the boundary obtained for increasing current. The hysteresis indicates that the transition between different phases has first order features.

\noindent {\it Discussion --} Using typical parameters, 
we estimate the current density at which the instabilities of skyrmion destruction occur. The current in Eq. \eqref{eq2} is given in units of $\gamma  e M_s^2\xi/{\mu_B}$, where $M_s\approx 10^3\ \rm{Gauss}$ is the saturation magnetization, $\gamma\approx 10^7\ \rm{Gauss^{-1}s^{-1}}$ is the gyromagnetic ratio \cite{Li04}, $\xi\approx 10\ \rm{nm}$ is the size of the skyrmion \cite{Yu2010a}, $e$ is the elementary charge, and $\mu_B$ is the Bohr magneton. We thus estimate the instability occurs at a current of order $10^{12}\ \rm{A/m^2}$, which is experimentally accessible. The fastest possible velocity for skyrmions is of the order of $\xi\gamma M_s\sim 100\ \rm{m/s}$ for a weak damping $\alpha\ll 1$ and it decreases as $1/\sqrt{1+\alpha^2\eta^2}$. To examine the instability and  motion of a single skyrmion in the FM background it should be possible to create a single skyrmion experimentally by applying a local magnetic field or a circular current \cite{Tchoe12}.

We emphasize that the dynamic phase transition at $J=J_m$ occurs regardless of the existence of a skyrmion in the original ($J=0$) FM state. The creation of skyrmions at $J_m$ can also be induced by other relaxation mechanisms, such as spin fluctuations induced by coupling to the lattice and conduction electrons. Such fluctuations provide the dominant relaxation mechanism far away from the moving skyrmion. The dynamical creation of skyrmions by current happens even at strong magnetic fields where the ground state is FM at $J=0$, which points to a possibly robust way to create skyrmions experimentally. Moreover, the resulting skyrmions form a liquid state, which has not yet been realized experimentally by only applying external magnetic fields. 

Finally, we  briefly discuss the effect of defects and compare the dynamic behavior of skyrmions  to that of vortices in type II superconductors. The pinning of skyrmions is weak as revealed by several experiments.~\cite{Jonietz2010,Yu2012,Schulz2012} Moreover, in the flow region,  the pinning potential is quickly averaged out by the fast moving skyrmions, similar to the case of  vortices. \cite{Koshelev94,Besseling03}. Thus, defects have a weak effect on the dynamical phase transition of skyrmions. Vortices in type II superconductors can also be created  by the external current, when the magnetic field induced by the current is larger than the lower critical field $H_{c1}$. For instance, vortex and anti-vortex pairs are created when  a current flows through a superconducting strip. For large currents, the normal core shrinks because the quasiparticles  are pumped out of the core by the electric field induced by the vortex motion \cite{Larkin65}. At a threshold current, the flux flow of the vortex lattice becomes unstable and the system jumps into the normal state. The dynamic creation of skyrmions has a completely different origin. It is induced by a chiral instability of the ferromagnetic state and no anti-skyrmion is created. The second instability of the skyrmion liquid is caused by a distortion produced by the damping.

\acknowledgments
We thank Christian Pfleiderer, Shinichiro Seki, Ivar Martin, Yasuyuki Kato, Leonardo Civale for useful discussions and Cynthia Reichhardt for a critical reading of the manuscript. Computer resources for numerical calculations were supported by the Institutional Computing Program in LANL.
This work was carried out under the auspices of the NNSA of the US DoE at LANL under Contract No. DE-AC52-06NA25396, and was supported by the US Department of Energy, Office of Basic Energy Sciences, Division of Materials Sciences and Engineering.


\begin{thebibliography}{37}
\bibitem [{not()}]{noteNumerics}%
  \BibitemOpen
  \href@noop {} {}\bibinfo {note} {In simulations, we take
  $J_{\rm{ex}}=D=\gamma=1$. The size of the simulation box is $L\times
  L=100\times 100$. The system is discretized with grid size $0.4$; a smaller
  grid size is also used to check the accuracy of the results. We use periodic
  boundary condition in both directions. To find the ground state, we anneal
  the system by adding a Gaussian noise field along the $z$ direction in
  $\mathbf{H}_{\rm{eff}}$. Equation \eqref{eq2} is solved by an explicit
  numerical scheme developed in Ref. \onlinecite{Serpico01}. The current is
  along the $x$ direction.}\BibitemShut {Stop}%
\end{thebibliography}
%

\end{document}